# Algebraic theory of an electro-optic modulator

G.P. Miroshnichenko and A. Gleim

Saint-Petersburg National Research University of Information Technologies, Mechanics and Optics, ITMO University, Kronverksky Avenue, 49,
197101 Saint Petersburg, Russia

Hamiltonian of a parametric process describing the interaction of a finite number of optical cavity modes, with the microwave field is proposed. Three-boson interaction is due to electro-optical effect. Analysis of the model is based on the algebraic properties of multimode field operators who, under certain assumptions, are su (2) algebra generators. The features of the transformation of the quantum states of signal photons in the electro-optical modulator are investigated. The limit on the number of interacting modes from the proposed (*restricted*) model to the traditionally used (*unrestricted*) model of electro-optical modulator is investigated.

## 1. Introduction

The classical theory of electro-optic effect is well developed [1, 2,3,4,5,6,7]. Such optical devices as modulators of light, returnable spectral filters, polarizing converters, optical communication circuits, and other devices operate on this effect. To describe a signal modulated by means of low-frequency periodic process on a frequency $\tilde{\Omega}$, researches use a Fourier series

$$E_{FM}(t) = A\exp\left(-i\left(\omega_{opt}t + \mu \cdot \cos\left(\tilde{\Omega}t\right)\right)\right) =$$
$$= A\exp\left(-i\omega_{opt}t\right)\sum_{n=-\infty}^{\infty}(-i)^n \exp\left(-in\tilde{\Omega}t\right)J_n(\mu) \quad (1)$$

Here $J_n(\mu)$ - the Bessel function of the first type of the integer index, $\mu$ - the modulation index, $\omega_{opt}$ - the frequency of carrier mode. Electro-optic phase modulator is widely used in quantum information processing technology to control quantum, single-photon and biphoton field states for quantum key distribution [8, 9, 10, 11, 12, 13, 14, 15] . In their works [16, 17] , the authors point out the necessity of developing the quantum theory of frequency and phase modulation. Quantum description of phase modulation requires a quantum description of phase. The early works [ 18, 19] bore the issue of quantum phase operator and phase measurement. The quantum theory of phase and instantaneous frequency, as well as interferometric methods of measurement are described in the works [16, 17]. Quantization of band-limited optical fields was developed in these studies. For this purpose, the authors identify the slowly varying envelope of creation operator and limit his spectrum of a narrow band around the carrier frequency. The authors use discrete time representation in their works with the help of Nyquist–Shannon sampling theorem instead of continuous time representation. The authors introduce the Fock space for the discrete modes, numbered with sample points. To reconstruct phase measurement procedure in the selected time interval the authors use the eigenfunctions of the phase operator and the positive definite operator measure which can be found in the works [20, 21]. In [22, 23] ,

unitary scattering matrix formalism is used to describe the transformation of the quantum state of the field in the electro-optic modulator. These works consider an electro-optic modulator as a multi-port beam splitter. The type of the scattering matrix is defined on the basis of the classical idea of the process of phase modulation due to the electro-optical effect in the nonlinear crystal. The scattering operator works in discrete Fock space based on a countable set of optical modes. Using the isomorphism of single-photon subspace and Hilbert space of a harmonic oscillator, the authors obtain the spectral decomposition of the scattering operator in the one-photon and multiphoton subspaces. Using the approach of band-limited optical field, the authors define the modulated wave packet obtained by passing the coherent quantum impulse through the modulator. As noted in these studies, the periodic sequence of the synchronized short pulses is a classical analogue of such a package. An alternative approach to the phase modulation was developed in the works [1- 7]. The method of the coupled classical modes of electromagnetic field is used in these works. According to the authors, it is the interaction of the eigenmodes of the cavity caused by the periodic modulation of the dielectric constant of the nonlinear crystal placed in the resonator leads to a phase modulation of the laser radiation. The laser impulse is a train of short emission impulses. The authors call this effect the internal phase modulation. Consistent quantum Hamiltonian theory of the electro-phase modulator using the idea of the mode interaction was developed in the work [24]. In this paper, the authors use the Hamiltonian of the interaction modes in single photon subspace and they take into account the dependence of the phase velocity of the optical and radio frequency waves on the frequency. The same system of differential-difference equations for the expansion coefficients $C_n(t)$ of modulated wave at basic modes is solved in the works [1 - 7] and in the work [24]

$$i\frac{d}{dt}C_n(t) + n\omega C_n(t) = \gamma\left(C_{n+1}(t) + C_{n-1}(t)\right),$$

$$C_k(t) = (-i)^n \exp\left(i\frac{\omega}{2}nt\right) J_n(\mu), \qquad (2)$$

$$\mu = \frac{4}{\omega}\gamma\sin\left(\frac{\omega}{2}T\right).$$

Where $\omega$ - is the out of sync frequency of the optical and radio waves caused by the dispersion of the phase velocity, $T$ - time interaction. Comparison of the formulas (1) and (2) shows that the distribution of the amplitudes of the optical modes, obtained from the equation system (2), provides an optical signal modulated in phase in the classic approach and in the quantum case. The works mentioned above presuppose that an infinite number of optical modes interact. Hereinafter such a model of an electro-optic modulator will be called *an unrestricted model*. The interaction parameter of two modes $\gamma$ does not depend on their number in this model. This assumption simplifies the analysis of an unrestricted model. The theory of this model in both the quantum and the classical case is based on a system of difference-differential equations (2).

Nowadays new schemes of optic modulator are well developed. A modulator, which is a micro-photonic resonator, combined with the microwave radio frequency resonator is studied in the works [25, 26, 27, 28, 29]. An optical resonator is a microdisk cut out of the crystal $LiNbO_3$. RF electrodes are disposed around the periphery of the disc. This device is effective because the optical mode "whispering gallery" is localized near the radiofrequency electrode, and it propagates in the electro-optic crystal in the presence of alternating high intensity microwave field. Such resonators have nonequidistant eigenmode spectrum. Thus, a small number of modes interact when the microwave field is turned on. The intensity of the sidebands experiences the saturation effect depending on the intensity of the microwave field. The modulator theory based on the three-mode interaction is described in the work [30].The semiclassical and the quantum theory of electro-optic modulator is a special case of the theory of parametric processes developed in the works [31, 32, 33]. Nowadays a new promising quantum information science discipline evolves. It appeared at the intersection of such disciplines as nanophysics and quantum light theory. This research area is called resonator optomechanics [34, 35, 36, 37], Further study of the parametric processes theory is required to analyze it. Thus, the researchers expect new progress in study of the Casimir effect, in the formulation of new protocols of quantum communication, quantum computing, quantum memory, and a host of other things.

The Hamiltonian of the parametric process, describing the interaction of a finite number of optical modes of the resonator containing a nonlinear crystal is studied in this paper. This model will be called *a restricted model* of a modulator. The effectiveness of the interaction caused by the effect of the microwave field on the resonator permittivity depends on the coefficient of the wave function overlapping of these modes. In reality, this coefficient is a function $f(k)$ o of mode numbers, and this limits the number of the interacting modes thus greatly complicating the study. This paper discusses the function $f(k)$, as a function satisfying the natural physical conditions. As a result of our assumption semiclassical Hamiltonian of optical modulator becomes an element of the algebra su(2) and analysis of the dynamic properties of the modulator is greatly simplified. Such choice of the optical mode overlapping function simplifies the quantum problem because the method of polynomial deformed quantum algebras can be used for the analysis [38].

This work has the following structure. Section 2 contains the description of the semiclassical Hamiltonian of the electro-optical modulator. Algebraic structure of the multimode operators used to build the model is analyzed in Section 3. In Section 4, we found the diagonal representation of the operator of quasi-energy. In section 5 we investigated the count rate of photons transmitted through the modulator, depending on the frequency. The passage to the limit to *unrestricted model* is studied Section 6. Here is the graphic matter showing specific features of the two models of modulator.

## 2. The Hamiltonian of the electro-optical modulator

An electro-optic modulator consists of a nonlinear crystal with the length $=L$, located between the metal electrodes. Radio frequency (microwave) wave is exited between the electrodes and propagates through the crystal from left to right. This wave due to electro-optic effect modulates the phase of the optical wave, changing its frequency content. Running optical cavity modes meet the periodic boundary conditions. The one-dimensional case sees that the wave vector of modes $k_n$ satisfies the equation

$$k_n L = 2\pi n, \quad n = \pm 1, \pm 2....., \quad k_n = -k_{-n} \ ..\tag{3}$$

Optical (carrier) mode excited by a laser impulse and has the frequency

$$\omega_{opt} = \omega_{\tilde{m}} = |k_{\tilde{m}}| v_{opt} = \Omega \cdot |\tilde{m}|, \quad \Omega = \frac{2\pi}{L} v_{opt}.$$

There $v_{opt}$ is the phase velocity of light in the used dielectric on the optical frequency. The number of the central working optical mode is designated as $\tilde{m}$. Usually value $\tilde{m}$ is the order $10^4 - 10^6$. The microwave radio frequency mode has the wave vector $k_{MW} = \frac{2\pi}{L}$ and the frequency $\tilde{\Omega} = k_{MW} v_{MW} = \Omega \frac{v_{MW}}{v_{opt}}$. There $v_{MW}$ is the phase velocity of the microwave mode.

Due to the electro-optic effect the basic modes interact on the segment $L$. Wave satisfying the conditions of conservation of energy and momentum, interact most effectively. The Hamiltonian of photons which takes into account these resonant summands has the form

$$H_0 = \tilde{\Omega} b^\dagger b + \Omega B_0 + \frac{\tilde{\gamma}}{f_{max}} \left( A^+ b + A^- b^\dagger \right), \quad (4)$$

$$B_0 = \sum_{m=1} m a_m^\dagger a_m, \quad A^- = \sum_{m=1} f(m) a_m^\dagger a_{m+1}, \quad A^+ = \sum_{m=1} f(m) a_m a_{m+1}^\dagger. \quad (5)$$

Where $a_n^\dagger, a_n, n = \pm 1, \pm 2...$ - the creation and annihilation operators of photons of optical modes with the number $n$, $b^\dagger, b$ - the creation and annihilation operators of photons of the microwave mode, $f(n)$ - function that takes into account the dependence of the force of modes interaction on their numbers. The coupling constant $\tilde{\gamma}$ of modes is normalized to the maximum value $f_{max}$ of the function $f(n)$. Nonresonant terms in (4) are discarded. For example, such terms are

$$a_{-n} a_{-(n+1)}^\dagger b, \quad a_n a_{-(n+1)}^\dagger b, \quad a_n^\dagger a_{n+1} b^\dagger, \quad n > 0.$$

These terms do not retain any momentum or energy of the photons thus these modes are weakly-coupled. Next, we ignore the quantum properties of microwave modes considering it as strongly exited. For this purpose, we replace the operators $b^\dagger, b$ with the c-numbers in (4) and obtain the Hamiltonian of the electro-optical modulator

$$H_0 = \Omega B_0 + \frac{\gamma}{f_{max}} \left( e^{-i\varphi} e^{-i\tilde{\Omega} t} A^+ + e^{i\varphi} e^{i\tilde{\Omega} t} A^- \right). \quad (6)$$

Where

$$\gamma = \tilde{\gamma} b. \quad (7)$$

## 3. The algebraic properties of the operators of optical modes

Real experiments show that a finite number of modes in the cavity interact. We introduce the parameters $n_{min}$, $n_{max}$, that make sense border set of numbers $m$ of the interacting modes

$$n_{min} < m \leq n_{max}. \quad (8)$$

These parameters $n_{min}$, $n_{max}$ may possess any positive integer. In accordance with this definition, we introduce the function $f(m)$

$$f(m) = \sqrt{(m - n_{min})(n_{max} - m)}.$$

The modes with the numbers outside the part (8) are not excited and are in a vacuum state. Thus, the operators (5) can be written over

$$B_0 = \sum_{m=n_{min}+1}^{n_{max}} m a_m^\dagger a_m, \quad A^- = \sum_{m=n_{min}+1}^{n_{max}-1} f(m) a_m^\dagger a_{m+1}, \quad A^+ = \sum_{m=n_{min}+1}^{n_{max}-1} f(m) a_m a_{m+1}^\dagger. \quad (9)$$

The mode with the number $\tilde{m}$

$$\tilde{m} = \frac{1 + n_{min} + n_{max}}{2}$$

has the frequency

$$\omega_{opt} = \tilde{m} \Omega.$$

Photons at this frequency excite the sidebands of the modulator. It is opportune to perform the summation shift

$$m = \tilde{m} + \Delta m, \ -S \leq \Delta m \leq S, \ S = \frac{n_{\max} - n_{\min} - 1}{2}.$$

Thus, the formulas (9) can be written over

$$B_0 = \tilde{m} \cdot N + A_0, \ A^- = \sum_{\Delta m = -S}^{S-1} f(\Delta m) a^\dagger_{\Delta m} a_{\Delta m + 1},$$

$$A^+ = \sum_{\Delta m = -S}^{S-1} f(\Delta m) a_{\Delta m} a^\dagger_{\Delta m + 1}, \quad A_0 = \sum_{\Delta m = -S}^{S} \Delta m \cdot a^\dagger_{\Delta m} a_{\Delta m}.$$

Here we introduce a new notation

$$f(\Delta m) = \sqrt{(S + 1 + \Delta m)(S - \Delta m)},$$

$$a_{\tilde{m} + \Delta m} \to a_{\Delta m}, \ a^\dagger_{\tilde{m} + \Delta m} \to a^\dagger_{\Delta m}, \ N = \sum_{\Delta m = -S}^{S} a^\dagger_{\Delta m} a_{\Delta m}. \quad (10)$$

The maximum value of the function $f(m)$ is equal to $\frac{2S+1}{2}$. Thus, the Hamiltonian operator (6) takes on this form

$$H_0 = \omega_{opt} \cdot N + \Omega A_0 + \frac{2\gamma}{2S+1}\left(e^{-i\varphi} e^{-i\tilde{\Omega}t} A^+ + e^{i\varphi} e^{i\tilde{\Omega}t} A^-\right). \quad (11)$$

The operators $A_0$, $A^+$, $A^-$ are the generators su(2) algebra, since they satisfy the commutation relations

$$\left[A_0, A^+\right] = A^+, \left[A_0, A^-\right] = -A^-, \left[A^+, A^-\right] = 2A_0. \quad (12)$$

The operator $K^2$ commutes with the generators. It is the Casimir operator

$$K^2 = A_0^2 + \frac{A^+ A^- + A^- A^+}{2}.$$

The operator $N$ also commutes with all the generators. Obviously, such a simple algebra arises due to the choice of the function $f(m)$ (10).

## 4. The quasi-energy spectrum of Hamiltonian

For the time-periodic Hamiltonian operator (11) we define the Hermitian operator of quasi-energy $Q$ and we find its quasi-energy spectrum. For this purpose, we present the evolution operator $U(t)$ of the Hamiltonian (11) in the form

$$U(t) = u(t)\exp(-iQt). \quad (13)$$

The evolution operator $U(t)$ satisfies the equation

$$i\frac{d}{dt}U(t) = H_0 U(t), \ U(0) = I.$$

The periodic operator $u(t)$ with the period $\frac{2\pi}{\tilde{\Omega}}$ produces the transition into the "rotating" coordinate system

$$u\left(t+\frac{2\pi}{\tilde{\Omega}}\right)=u(t)=\exp\left(-i(\tilde{m}N+A_0)\tilde{\Omega}t\right),$$

$$\varphi=0.$$

As a result of the manipulation we obtain the quasi-energy operator (Hamiltonian in the "rotating" coordinate system)

$$Q=\omega(\tilde{m}N+A_0)+\frac{2\gamma}{2S+1}(A^++A^-),$$

$$\omega=\Omega-\tilde{\Omega}.$$

The operator $Q$ can be diagonalized by a unitary transformation $U_0$

$$U_0 Q U_0^\dagger = 2\Gamma\cdot A_0 + \omega\tilde{m}N, \quad U_0=\exp(-i\beta\cdot\Phi), \quad \Phi=i(A^+-A^-),$$

$$\sin(2\beta)=\frac{2\gamma}{(2S+1)\Gamma}, \quad \Gamma=\sqrt{\left(\frac{\omega}{2}\right)^2+\left(\frac{2\gamma}{2S+1}\right)^2}.$$
(14)

The operator $Q$ has equidistant degenerate spectrum, the distance between the levels is $\Gamma$. The eigenvectors differ by a set of quantum numbers: $|A_0,K,N,\xi\rangle$. Where $K$ - the eigenvalue of the Casimir operator, $A_0$ - eigenvalue (projection) of the operator $A_0$, $-K\leq A_0\leq K$, $N$ is the eigenvalue of the number of photons $N$ (10), $\xi$ - the value differentiating the basis vectors with the same projection $A_0$ and different value $K$. The set of the basis vectors with the fixed $N$ has the dimension $\frac{(N+2S)!}{(2S)!(N)!}$. The maximum value of $K$ is equal to $K=K_{max}=N\cdot S$.

The values of the other numbers $K$ in the set can be determined as the case may be.

## 5. Counting rate of photons at the modulator output

Let us suppose that modulator cavity modes are excited by photons of the carrier frequency running through the device from left to right. A traveling wave of microwave field extends inside the cavity. We denote the modes interaction time in the resonator as $T$ and calculate the mode density matrix at the moment $T$. If we neglect the losses, we obtain

$$\rho_a(T)=U(T)\rho_a(0)U^\dagger(T).$$
(15)

Where $U(T)$ - the evolution operator (13). We calculate the average number of photons in the mode with the number $\Delta m$ at the time $T$

$$\bar{n}_{\Delta m}(T)=Sp_a\left(a^\dagger_{\Delta m}a_{\Delta m}\rho_a(T)\right)=Sp_a\left(a^\dagger_{\Delta m}(T)a_{\Delta m}(T)\rho_a(0)\right).$$

We recognize

$$a_{\Delta m}(T)=\exp(iQT)u(T)^\dagger a_{\Delta m}u(T)\exp(-iQT)=$$
$$=\exp\left(-i(\tilde{m}+\Delta m)\tilde{\Omega}T\right)\exp(iQT)a_{\Delta m}\exp(-iQT)$$
(16)

We use (14), and get

$$\exp(iQT)a_{\Delta m}\exp(-iQT)=$$
$$=U_0^\dagger\exp\left(i(2\Gamma\cdot A_0+\omega\tilde{m}N)T\right)U_0 a_{\Delta m}U_0^\dagger\exp\left(-i(2\Gamma\cdot A_0+\omega\tilde{m}N)T\right)U_0.$$

The transformation of the operators $a_{\Delta m}$, $a^\dagger_{\Delta m}$ is linear. We put this transformation with the matrix $D$

$$U_0 a_{\Delta m} U_0^\dagger = \sum_{\Delta k=-S}^{S} D_{\Delta m,\Delta k} a_{\Delta k}.$$

We use the Baker–Campbell–Hausdorff formula and the formul (14)

$$U_0 a_{\Delta m} U_0^\dagger = \exp(-i\beta\cdot\Phi) a_m \exp(i\beta\cdot\Phi) =$$

$$= \sum_{p=0}^{\infty} \frac{(-i\beta)^p}{p!} \left[\Phi,\left[\Phi,\ldots\left[\Phi, a_{\Delta m}\right]_p \ldots\right]_2\right]_1 = \sum_{\Delta k=-S}^{S} D_{\Delta m,\Delta k} a_{\Delta k}.$$

The operator is the following

$$[\Phi, a_{\Delta m}] = \sum_{\Delta k=-S}^{S} F_{\Delta m,\Delta k} a_{\Delta k},$$

$$F_{\Delta m,\Delta k} = i\left(\delta_{\Delta m,\Delta k-1}\cdot f(\Delta k-1) - \delta_{\Delta m,\Delta k+1}\cdot f(\Delta k)\right).$$

Where $\delta_{m,k}$ - the Kronecker delta. The matrix $D$ with the dimensionality $2S+1\times 2S+1$ has the form

$$D = \exp(-i\beta F).$$

It is convenient to express the matrix $F$ in terms of the spin matrix $S_y$ in spin space for the spin $S = \dfrac{n_{max}-n_{min}-1}{2}$. Thus we obtain

$$F = 2S_y.$$

The matrix $D$ - is d-Wigner function of the argument $2\beta$ [39]

$$D_{\Delta m,\Delta k} = d^S_{\Delta m,\Delta k}(2\beta), \quad D^{-1} = D^T, \quad (17)$$
$$-S \leq \Delta m, \Delta k \leq S.$$

The operator (16) has the form

$$a_{\Delta m}(T) = \exp\left(-i\left((\tilde{m}+\Delta m)\tilde{\Omega}+\omega\tilde{m}\right)T\right) \sum_{\Delta p=-S}^{S} a_{\Delta p} R_{\Delta m,\Delta p}(T),$$

$$R_{\Delta m,\Delta p}(T) = \sum_{\Delta k=-S}^{S} d^S_{\Delta m,\Delta k}(2\beta) d^S_{\Delta p,\Delta k}(2\beta) \exp(-i\Delta k\cdot 2\Gamma T).$$

We simplify the form of the matrix $R_{\Delta m,\Delta p}(T)$, using the d- Wigner functions addition formula

$$R_{\Delta m,\Delta p}(T) = \exp(-i\alpha(\Delta m+\Delta p))\cdot(-1)^{\Delta p}\cdot d^S_{\Delta m,\Delta p}(2\tilde{\beta}). \quad (18)$$

Here, the angles $\alpha$ and $2\tilde{\beta}$ can be evaluated according to [39]

$$\operatorname{ctg}(\alpha) = \frac{\cos(2\beta)}{\sin(2\Gamma T)}(\cos(2\beta)-\cos(2\Gamma T)),$$
$$\sin(2\tilde{\beta}) = 2\sin(2\beta)\sin(\Gamma T)\sqrt{1-(\sin(2\beta)\sin(\Gamma T))^2}. \quad (19)$$

Let us suppose that the initial matrix $\rho_a(0)$ such that only the central mode is excited at the initial moment in the resonator $\Delta m = 0$. Thus we obtain

$$\bar{n}_{\Delta m}(T) = |R_{\Delta m,0}(T)|^2 \bar{n}_{\Delta m}(0),$$
$$\bar{n}_{\Delta m}(0) = Sp_a\left(a_0^\dagger a_0 \rho_a(0)\right). \qquad (20)$$

The output of the resonator is connected to the filter Fabry - Perot via fiber channel. A photodetector with a sufficiently wide bandwidth is located after of filter. It is necessary to match the resonator modes and the modes of the optical fiber. Let us suppose that they are correlated perfectly

$$a_{\Delta m} = b_{\Delta m}, \; -S \leq \Delta m \leq S.$$

There $b_{\Delta m}$, $b_{\Delta m}^\dagger$ - the operators of the fiber modes. We image the photodetector by an atom and we denote the bound ground state by $|0\rangle_A$, we denote the excited state of the continuous spectrum by $|\varepsilon\rangle_A$, we denote the frequency of boundary of the continuous spectrum by $\omega_0$, the Hamiltonian operator of the atom and the fiber modes interaction is $V_{bA}$

$$V_{bA} = V_{bA}^+ + V_{bA}^-.$$

Where

$$V_{bA}^+ = E^+ \cdot d^-, \; V_{bA}^- = \left(V_{bA}^+\right)^\dagger,$$

$$d^- = \sum_\varepsilon \delta_\varepsilon |\varepsilon\rangle_A \langle 0|, \; E^+ = \sum_{\Delta m=-S}^{S} \phi_{\Delta m} b_{\Delta m},$$

$\delta_\varepsilon$ - transition dipole moment in the continuous spectrum, $\phi_{\Delta m}$ - parameter depending on the frequency of the mode and of the mode wave function at the location of the atom detector. According to the Mandel's theory [40], the photon counting rate $p_0(\omega_f)$ can be simply calculated in the second order for the operator $V_{bA}$

$$p_0(\omega_f) = \lim_{t\to\infty} \int_{-t}^{t} W(\tau) d\tau. \qquad (21)$$

The correlation function $W(\tau)$ has the form

$$W(\tau) = |\delta_{\omega_{opt}}|^2 g(\omega_{opt}) \sum_{\Delta m=-S}^{S} \bar{n}_{\Delta m}(T) \exp(-i\Omega(\tilde{m}+\Delta m)\tau) K(\tau, \omega_f). \qquad (22)$$

Here we have used the Markov approximation [41] in the transition from the summation of the atomic spectrum to integration

$$\sum_\varepsilon |\delta_{\omega_\varepsilon}|^2 \exp(i\omega_\varepsilon \tau) K(\omega_\varepsilon, \omega_f) =$$
$$= \frac{1}{2\pi} \int_{-\infty}^{\infty} \exp(i\omega_\varepsilon \tau) |\delta_{\omega_\varepsilon}|^2 g(\omega_\varepsilon) K(\omega_\varepsilon, \omega_f) d\omega_\varepsilon \approx |\delta_{\omega_{opt}}|^2 g(\omega_{opt}) K(\tau, \omega_f). \qquad (23)$$

It is assumed that there is the frequency filter with a spectral transmittance bandwidth $K(\omega, \omega_f)$ before the detector and the value of the bandwidth of the frequency filter is smaller than the modulation frequency $\Omega$. Where $\omega_\varepsilon$ - the atom transition frequency from the ground state to the continuous spectrum, $\omega_f$ - the centre frequency of the filter band pass. In the formula (23) the lower limit of integration is $-\infty$, as the spectral density of the atomic states satisfies the equation

$$g(\omega_\varepsilon) = 0, \; \omega_\varepsilon \leq \omega_0.$$

The formula (23) presupposes that $|\delta_{\omega_\varepsilon}|^2$, $g(\omega_\varepsilon)$ change in a small way in the filter band pass that is why these factors can be taken out of the integral. We substitute (20), (22), (23) in (21), and obtain the formula for the photon rate counting depending on the filter tuning and on the modes interaction time $T$ in the modulator

$$p_0(\omega_f, T) = p_0(\omega_{opt}, 0) p_{rel}(\omega_f, T),$$

$$p_{rel}(\omega_f, T) = \sum_{\Delta m = -S}^{S} |R_{\Delta m, 0}(T)|^2 K(\omega_{opt} + \Omega \Delta m, \omega_f), \quad (24)$$

$$p_0(\omega_{opt}, 0) = |\delta_{\omega_{opt}}|^2 g(\omega_{opt}) Sp_a(a_0^\dagger a_0 \rho_a(0)) |\phi_0|^2.$$

The time $T$ determines the value of the modulation index. Where $p_0(\omega_{opt}, 0)$ - counting rate at the optical frequency when modulator is switched off. The figures (fig.1-5) show the diagrams of the velocity ratio of photon counting $p_{rel}(\omega_f, T)$.

## 6. Results and conclusion

The formulas (15), (24) show how the quantum state and the spectrum of the optical mode signal is converted by the action of the electro-optical modulator. The finite number of the cavity modes with numbers $\Delta m$ belonging to the interval $-S \leq \Delta m \leq S$ covered parametric interaction. We compare our results with those found for the *unrestricted* model in the classical limit [5], and in the quantum description [22, 23, 24]. For this purpose we take the limit $S \to \infty$ in the above formulas. We write the d-Wigner function (17) by means of the Jacobi polynomial [39]

$$d_{\Delta m, \Delta k}^S(2\beta) = \xi_{\Delta m, \Delta k} \sqrt{\frac{s!(s+\mu+\nu)!}{(s+\mu)!(s+\nu)!}} \cdot (\sin\beta)^\mu \cdot (\cos\beta)^\nu \cdot P_s^{(\mu,\nu)}(\cos 2\beta), \quad (25)$$

$$\mu = |\Delta m - \Delta k|, \quad \nu = |\Delta m + \Delta k|, \quad s = S - \frac{1}{2}(\mu + \nu).$$

According to [42], the asymptotic of the Jacobi polynomial has the form

$$\lim_{n \to \infty} \left[ n^{-\alpha} \cdot P_n^{(\alpha,\beta)}\left(\cos\frac{z}{n}\right) \right] = \lim_{n \to \infty} \left[ n^{-\alpha} \cdot P_n^{(\alpha,\beta)}\left(1 - \frac{z^2}{2n^2}\right) \right] = \left(\frac{z}{2}\right)^{-\alpha} J_\alpha(z). \quad (26)$$

We obtain the asymptotic formulas for the angles $\alpha$ and $2\tilde{\beta}$ with the help of (19)

$$\alpha \xrightarrow{S \to \infty} \frac{\pi - \omega T}{2},$$

$$S \cdot 2\tilde{\beta} \xrightarrow{S \to \infty} \frac{4\gamma}{\omega} \sin\left(\frac{\omega T}{2}\right).$$

The asymptotic behavior of the matrix $R_{\Delta m, \Delta p}(T)$ (18) is achieved with the help of the formulas (25), (26)

$$R_{\Delta m, \Delta p}(T) \xrightarrow{S \to \infty} (-i)^{(\Delta p + \Delta m)} \cdot \exp\left(i(\Delta p + \Delta m)\frac{\omega}{2}T\right) \cdot \exp(-i\pi \Delta p) \cdot J_{\Delta m - \Delta p}(\mu). \quad (27)$$

Here is the modulation index

$$\mu = 4\frac{\gamma}{\omega}\sin\left(\frac{\omega}{2}T\right). \tag{28}$$

This formula correlates with the formula (2). The resulting limit formulas (27), (28) are correct under the condition

$$|\Delta m|, |\Delta k| \ll S, \quad \omega \neq 0.$$

This result coincides with the corresponding results of the works [22, 23, 24]. We find the average value of the operator of the electric field at the output of the electro-optical modulator in the limit $S \to \infty$

$$\bar{E}^+ = Sp_a\left(\rho_a(T)\sum_{\Delta m=-S}^{S} E_{\Delta m} a_{\Delta m}\right) \approx E_0 \sum_{\Delta m=-S}^{S} Sp(\rho_a(0)a_{\Delta m}(T)) =$$

$$= E_0 Sp(\rho_a(0)a_0)\exp(-i\omega_{opt}T)\sum_{\Delta m=-S}^{S}\exp(-i\Delta m\tilde{\Omega}T)R_{\Delta m,0}(T) \xrightarrow[S\to\infty]{}$$

$$\xrightarrow[S\to\infty]{} E_0 Sp(\rho_a(0)a_0)\exp(-i\omega_{opt}T)\exp\left(-i\mu\cos\left(\left(\tilde{\Omega}-\frac{\omega}{2}\right)T\right)\right).$$

This formula corresponds to the classical one (1).

The dependence of the relative photon counting rate on the tuning frequency of the interference filter is shown in Fig. 1-3. The calculation is performed by the formula (24). Zero on the frequency axis corresponds to the filter tuning frequency of the optical carrier $\omega_f = \omega_{opt}$. Along the frequency axis, we will choose the relative scale. One unit of frequency scale corresponds to $1/30$ of the frequency $\Omega$ - the frequency difference between adjacent optical modes. The transmission coefficient of the frequency filter $K(\omega_{opt}+\Omega\Delta m, \omega_f)$ is imaged with the Gaussian curve with the half-width $\Delta\omega = 4$ at the height $1/e$. The calculations shown in Fig.1-5 are performed for the following parameters:

$$S = 3, \quad \omega = 0.1, \quad \Omega = 30, \quad T = \frac{2\pi}{\Omega} = 0.209.$$

Interaction parameter $\gamma$, which has the dimension of frequency, equal to: for Fig.1 - $\gamma_1 = 2$, for Fig.2 - $\gamma_2 = 10$, for Fig.3 - $\gamma_3 = 24.25$. There are 2 graphs in each figure: Graph 1 is designed for *a restricted* model, Graph 2 is designed for *an unrestricted* modulator model. These figures show a different behavior of the two models, depending on the frequency of the filter settings and the interaction parameter (7). At the initial time, the photons belong to the central mode. The figures show that with the increase of the interaction parameter (and an increase in modulation index) in the *restricted* model photons photons occupy a finite number of modes (the number of modes in the calculation is $2S+1 = 7$). There is no saturation effect on the parameter $\gamma$ in the *unrestricted* model, thus the photons "spread out" over a large (infinite in the limit) number of modes. This tendency is shown in Fig. 4 and 5 that is with increase of $\gamma$ the number of photons in the modes decreases. Whereas for the restricted model (Graph 1 in Fig. 4.5) number of photons oscillate around an average value that is independent of $\gamma$. We analyze the case $\omega = 0$ (the absence of out of sync phase velocities). In this case $2\beta = \frac{\pi}{2}$. For the interaction parameter $\gamma = \frac{\Omega}{8}(2S+1)$ with $T = \frac{2\pi}{\Omega}$, we obtain from the formula (19) $2\tilde{\beta} = \pi$. Using the property of the d-Wigner function [39]

$$d^S_{\Delta m, \Delta k}(\pi) = (-1)^{S+\Delta m} \cdot \delta_{\Delta m, -\Delta k},$$

we get

$$\left|R_{\Delta m, 0}(T)\right|^2 = \delta_{\Delta m, 0}.$$

That means that with $\omega = 0$ when $\gamma = \frac{\Omega}{8}(2S+1)$, $T = \frac{2\pi}{\Omega}$ the photons are collected in the central mode. The dependence of counting rate of photons in the central mode $\Delta m = 0$ on the interaction parameter $\gamma$ is shown in Fig. 4, and in Figure 5 for $\Delta m = 2$. The photon counting rate decreases with increase of $\gamma$ (for the *unrestricted* model, Graph 2). The restricted model (Graph 1) shows the revival effect for $\gamma = \frac{\Omega}{8}(2S+1)$ and $\gamma = \frac{\Omega}{4}(2S+1)$.

In conclusion, we may say that we analyzed the exactly soluble (*restricted*) model of the electro-optic modulator. The analysis of the model is based on the algebraic properties of multimode field operators $A_0$, $A^+$, $A^-$ (12), which are the algebra su(2) generators. The proposed Hamiltonian conserves the number of photons that are distributed between a finite numbers of optical modes. The total number of interacting optical modes is equal to $2S+1$. We investigated the limiting transition to the unlimited number of the interacting modes $S \to \infty$ and showed that many of the properties of our model in the limit coincide with those of the *unrestricted* model. The *unrestricted* model was examined in both the classical and quantum cases, in the works [1 - 7, 22, 23, 24]. A distinctive feature of the *unrestricted* model is the lack of saturation effect on the microwave field density. The *restricted* model has the revival effects, when some values of the rf field photons return to the central mode.

## Acknowledgments

The work was partially financially supported by the Government of the Russian Federation (grant 074-U01), by the Ministry of Science and Education of the Russian Federation (GOSZADANIE 2014/190, Project 14.Z50.31.0031), by grant of Russian Foundation for Basic Researches and grants of the President of Russia (state contracts 14.124.13.2045-MK and 14.124.13.1493-MK).

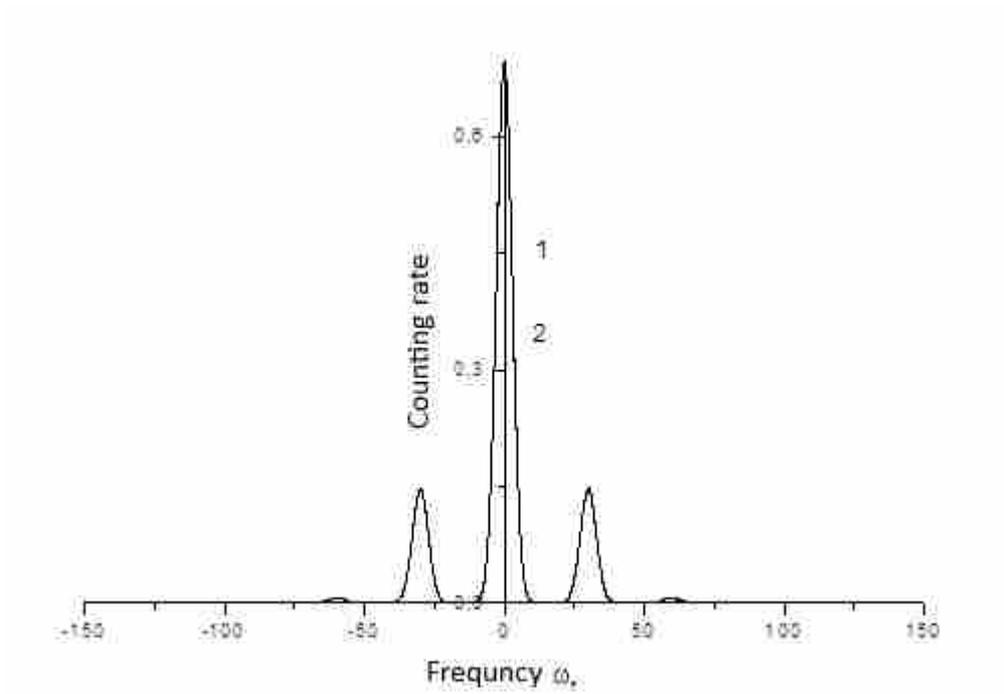

Fig.1. The relative counting rate of the photon, depending on the filter frequency tuning. The interaction parameter $\gamma = 2$. The calculation parameters, as well as the choice of scale of frequency and time are explained in this text. Two Graphs: 1 and 2 built for the *restricted* and *unrestricted* models do not differ much.

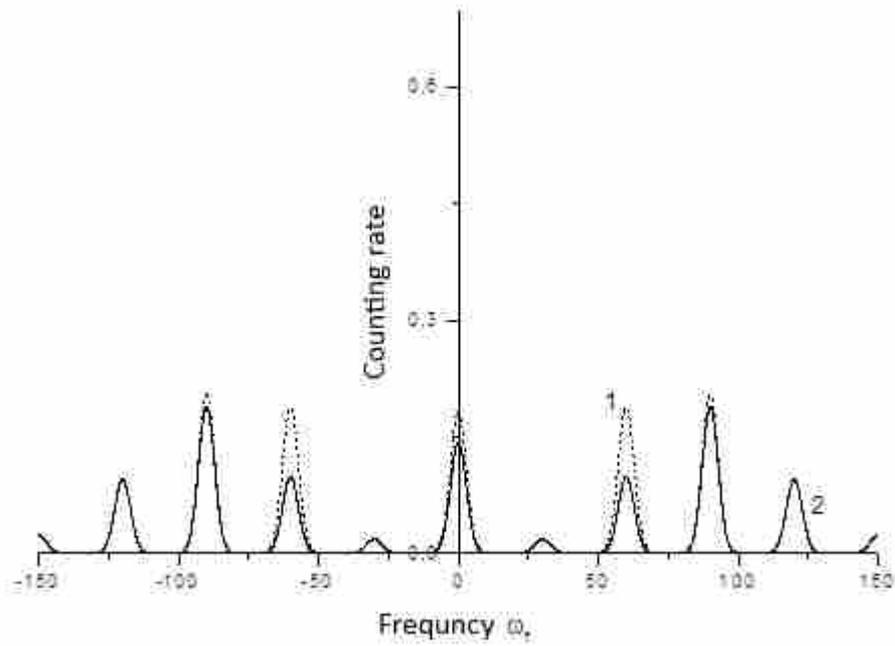

Fig.2. The relative counting rate of the photon, depending on the filter frequency tuning. The interaction parameter $\gamma = 10$. The calculation parameters, as well as the choice of scale of frequency and time are explained in this text. Graph 1 is for *restricted* model; Graph 2 is for *unrestricted* model. Photons "spread out" over the adjacent modes.

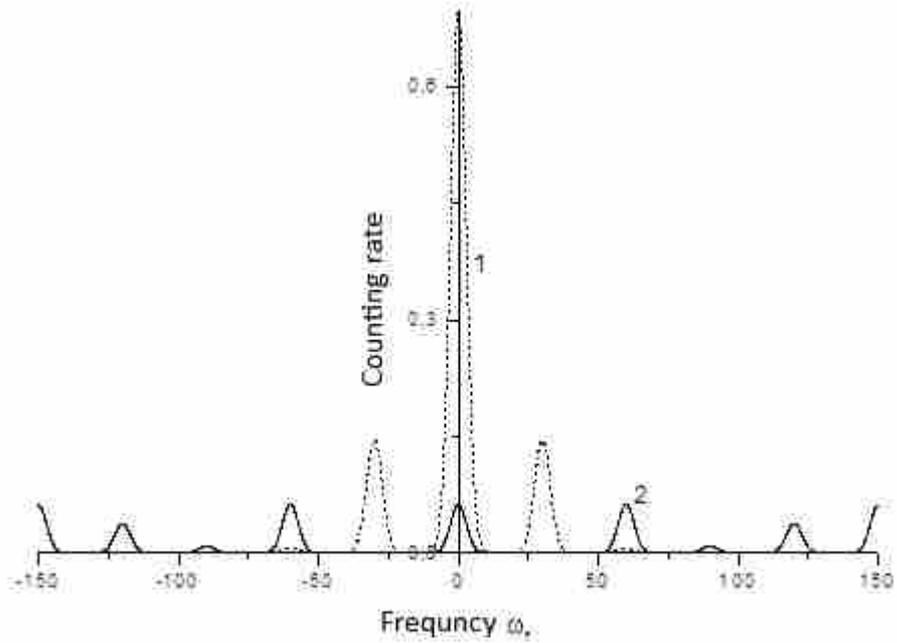

Fig.3. The relative counting rate of the photon, depending on the filter frequency tuning. The interaction parameter $\gamma = 24.25$. The calculation parameters, as well as the choice of scale of frequency and time are explained in this text. Graph 1 is for *restricted* model; Graph 2 is for *unrestricted* model. Graph 1 shows the effect of the revival of the photons in the central mode for *restricted* model.

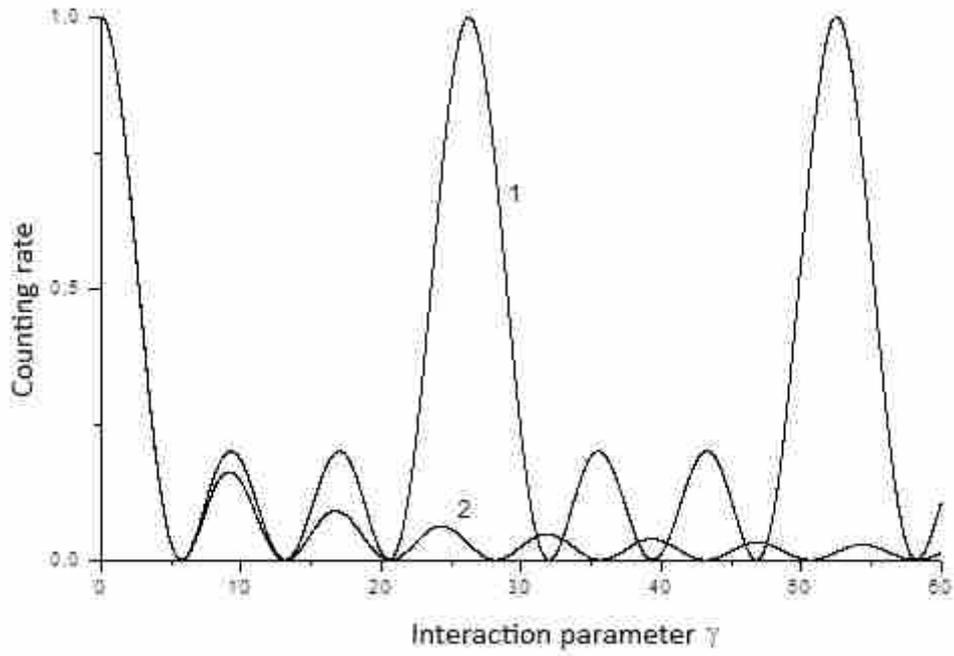

Fig.4. The relative counting rate of the photon in the central mode $\Delta m = 0$ depending on the interaction parameter. The calculation parameters, as well as the choice of scale of frequency and time are explained in this text. Graph 1 is for *restricted* model; Graph 2 is for *unrestricted* model.

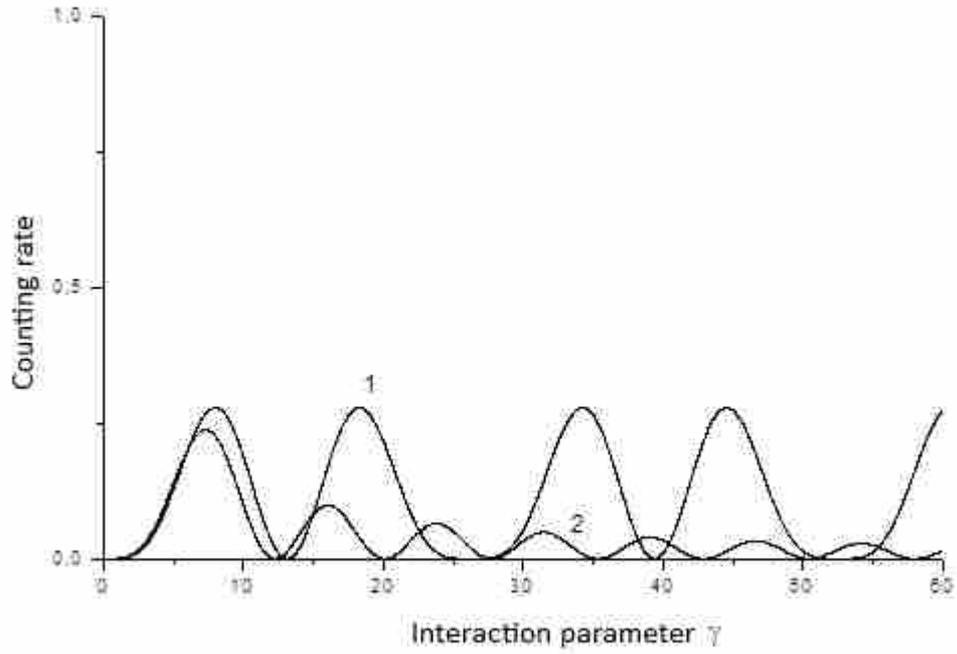

Fig.5. The relative counting rate of the photon in the central mode $\Delta m = 2$ depending on the interaction parameter. The calculation parameters, as well as the choice of scale of frequency and time are explained in this text. Graph 1 is for *restricted* model; Graph 2 is for *unrestricted* model.

# Abstract

Hamiltonian of a parametric process describing the interaction of a finite number of optical cavity modes, with the microwave field is proposed. Three-boson interaction is due to electro-optical effect. Analysis of the model is based on the algebraic properties of multimode field operators who, under certain assumptions, are su (2) algebra generators. The features of the transformation of the quantum states of signal photons in the electro-optical modulator are investigated. The limit on the number of interacting modes from the proposed (*restricted*) model to the traditionally used (*unrestricted*) model of electro-optical modulator is investigated.